\documentclass[aps,prl,reprint,groupedaddress,amsmath,amssymb]{revtex4-1}

\usepackage[dvips]{graphicx}
\usepackage{bm}

\begin{document}


\title{Magnetic Exciton Mediated Superconductivity\\ in the Hidden-Order Phase of URu$_{2}$Si$_{2}$}


\author{Hiroaki Kusunose}
\affiliation{Department of Physics, Ehime University, 790-8577, Matsuyama, Japan}


\date{\today}

\begin{abstract}
We propose the magnetic exciton mediated superconductivity occurring in the enigmatic hidden-order phase of URu$_{2}$Si$_{2}$.
The characteristic of the massive collective excitation observed only in the hidden-order phase is well reproduced by the antiferro hexadecapole ordering model as the trace of the dispersive crystalline-electric-field excitation.
The disappearance of the superconductivity in the high-pressure antiferro magnetic phase can naturally be understood by the sudden suppression of the magnetic-exciton intensity.
The analysis of the momentum dependence of the magnetic-exciton mode leads to the exotic chiral $d$-wave singlet pairing in the $E_{g}$ symmetry.
The Ising-like magnetic-field response of the mode yields the strong anisotropy observed in the upper critical field even for the rather isotropic 3-dimensional Fermi surfaces of this compound.
\end{abstract}


\maketitle

Itinerant-localized duality of $f$ electron in condensed matter often exhibits an enigmatic phase, whose order parameter is almost inaccessible by conventional experimental techniques \cite{Kuramoto09}.
One famous example is given in the heavy-fermion superconductor, URu$_{2}$Si$_{2}$, where the ordered phase signaled by the clear specific-heat anomaly at $T_{0}=17.5$ K has not been resolved over the past 25 years \cite{Mydosh11}.
The magnetic responses of this body-centered tetragonal compound are characterized by strong Ising anisotropy, and the type-I antiferro magnetic (AFM) phase with ${\bm Q}=(1,0,0)$ emerges in application of pressure through the 1st-order transition from the hidden-order (HO) phase \cite{Aoki09}.

Interestingly, the bulk superconductivity below $T_{\rm c}=1.5$ K coexists only with the HO phase \cite{Hassinger08}, although similarity of the Fermi surfaces in both the HO and the AFM phases is pointed out by Shubnikov-de Haas measurement \cite{Hassinger10}.
Moreover, the upper critical field, $H_{c2}(T)$ shows strong anisotropy, {\it i.e.}, $H_{c2}(0)$ for $H\,\|\,[001]$ is four times smaller than for $H\,\|\,[100]$, although the 3-dimensional Fermi surfaces are rather isotropic \cite{Ohkuni99}.
This anisotropy is not simply ascribed to the paramagnetic effect due to the Ising anisotropy in $\chi(T)$ \cite{Brison94}, since the NMR Knight shift measurement suggests that the dominant contribution of the susceptibility is the orbital part and the quasiparticle susceptibility is small and rather isotropic \cite{Matsuda96,Tou05}.

In the superconducting phase, the power-law $T$ dependences of the specific heat, $C(T)\propto T^{2}$ \cite{Hasselbach93}, and the NMR relaxation rate, $T_{1}^{-1}\propto T^{3}$ \cite{Kohori96} at low temperatures are consistent with the linear density of states at low-energy, suggesting the presence of a line node in the gap structure.
The recent thermal transport \cite{Kasahara07,Kasahara09} and the heat capacity \cite{Yano08} measurements in rotating magnetic fields find the multi-gap behavior and the additional point nodes along the c-axis, from which the chiral $d$-wave pairing, $k_{z}(k_{x}+ik_{y})$, in the $E_{g}$ symmetry has been proposed.

Recently, the antiferro hexadecapole (AFH) order of the U ion has been put forward for the HO phase based on the state-of-the-art numerical computation \cite{Haule09}.
The hexadecapole scenario by the Ginzburg-Landau analysis \cite{Haule10} and the localized multipole exchange model \cite{Kusunose11} have provided a natural and coherent description for numerous observations.
Especially, it provides a convincing explanation of the collective magnetic excitation observed in neutron scattering \cite{Broholm87,Broholm91}.
The mode at ${\bm Q}$ is observed only in the HO phase, but not in the AFM phase \cite{Villaume08} as a consequence of the strong Ising anisotropy.

By these circumstances, we immediately suspect that the superconductivity is mediated by the massive collective magnetic excitation --- magnetic exciton.
Its realization has been addressed so far in UPd$_{2}$Al$_{3}$ \cite{Sato01,Miyake01} and PrOs$_{4}$Sb$_{12}$ \cite{Matsumoto04}.
In this Letter, we first reproduce the characteristic of the observed magnetic-exciton mode on the basis of the AFH ordering model.
Then, we discuss which pairing symmetry is favored by the irreducible-symmetry decomposition of the momentum dependence of the mode.
We demonstrate that the strong anisotropy in $H_{c2}$ can be interpreted as the Ising-like magnetic-filed response of the magnetic-exciton intensity in a coherent manner with the AFH-AFM orders.

Let us begin with the multipole exchange model, 
\begin{multline}
H_{f}=-\sum_{\bm q}\gamma({\bm q})\left[J(p)\sigma_{\bm q}\sigma_{-{\bm q}}+D(p)\xi_{\bm q}\xi_{-{\bm q}}\right]
\\
-H\sum_{i}\sigma_{i}+\sum_{i}H_{i}^{\rm CEF},
\label{exmodel}
\end{multline}
where $\sigma$ and $\xi$ represent the multipoles proportional to the Ising dipole and the $xy(x^{2}-y^{2}$)-type electric hexadecapole respectively, being active in the assumed low-lying $\Gamma_{1}^{(1)}$-$\Gamma_{2}$-$\Gamma_{1}^{(2)}$ (0-50-170 K) crystalline-electric-field (CEF) scheme in the last term of (\ref{exmodel}) \cite{Kusunose11}.
$H$ is the magnetic field along the c-axis.
Note that $J_{x}$ and $J_{y}$ are inactive within this CEF scheme.
We assume the linear pressure dependences in $J(p)$ and $D(p)$, and the common momentum dependence $\gamma({\bm q})$, which is normalized as $\gamma({\bm Q})=1$.
Although in the previous study the nearest-neighbor interaction was used in the mean-field (MF) analysis for simplicity \cite{Kusunose11}, here the overall exchange couplings $J$ and $D$, and $\gamma({\bm q})$ have been determined to reproduce the observed phase diagram and the inelastic neutron scattering data.
To do so, we have required the rather long-range exchange interaction up to 7th neighbor,
\begin{multline}
\gamma({\bm q})=2j_{1}(c_{2x}+c_{2y})+8j_{2}c_{x}c_{y}c_{z}+4j_{3}c_{2x}c_{2y}
\\
+16j_{4}c_{x}c_{y}c_{z}(c_{2x}+c_{2y}-1)+2j_{5}(c_{4x}+c_{4y})
\\
+4j_{6}(c_{4x}c_{2y}+c_{2x}c_{4y})+2j_{7}c_{2z},
\end{multline}
where $c_{n\alpha}=\cos(n\pi q_{\alpha})$ with $\alpha=x,y,z$.

\begin{figure}[tb]
\includegraphics[width=8.5cm]{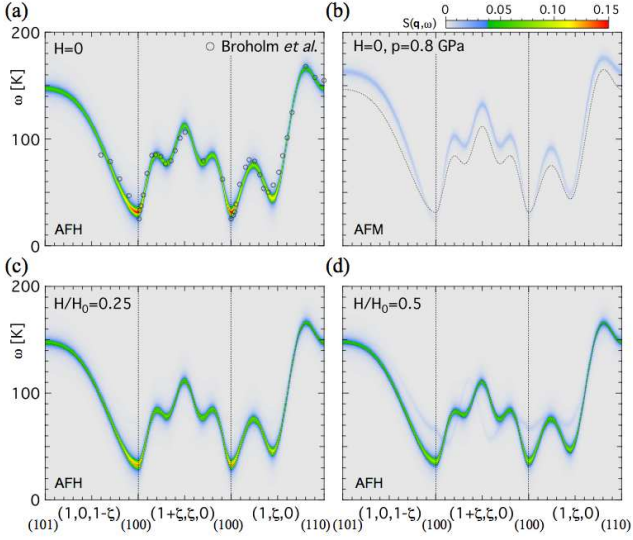}
\caption{(color online) The calculated magnetic excitation spectra at (a) ambient pressure. The symbols are taken from \cite{Broholm91}, (b) $p=0.8$ GPa, (c) $H/H_{0}=0.25$, where $H_{0}$ is the critical field of the AFH phase, and (d) $H/H_{0}=0.5$. The strongest intensity at ${\bm Q}$ in the AFH phase (a) disappears in the AFM phase (b). The dashed line indicates the dispersion in (a). The increase of $H\,\|\,[001]$ continuously decreases the intensity at ${\bm Q}$ as in (c) and (d). The spectra in the high-$H$ paramagnetic phase (not shown) is similar to (b) in the AFH phase.}
\label{skw}
\end{figure}

The longitudinal dynamical spin susceptibility of f-electron,
\begin{equation}
\chi_{z}({\bm q},\omega)=i\int_{0}^{\infty}dt\langle[\delta\sigma_{\bm q}(t),\delta\sigma_{-{\bm q}}(0)]\rangle e^{i\omega t},
\end{equation}
with $\delta\sigma_{\bm q}=\sigma_{\bm q}-\langle\sigma_{\bm q}\rangle$ can be calculated by using the Holstein-Primakoff method at $T=0$ \cite{Kusunose01,Shiina03}.
Then, the structure function, $S({\bm q},\omega)={\rm Im}\,\chi_{z}({\bm q},\omega)/\pi$, can be compared with the observed magnetic-exciton mode in the HO phase.
Figure \ref{skw}(a) shows the result of the fitting of the inelastic neutron scattering data \cite{Broholm91}.
The overall dispersion and the characteristic of the intensity peaked at ${\bm Q}$ and the incommensurate vector ${\bm Q}^{*}=(1,0.4,0)$ are well reproduced.
The present description of the magnetic excitation spectra provides an interpretation that the magnetic exciton is the trace of the CEF excitation which propagates by the exchange couplings.

In the application of the pressure, which increases slightly the overall exchange couplings, the property of the spectra is gradually changed, and the strongest intensity at ${\bm Q}$ in the AFH phase suddenly disappears upon entering the AFM phase.
As was discussed in \cite{Haule10,Kusunose11}, the suppression of the intensity at ${\bm Q}$ can naturally be understood by the change of the role for the transverse component at the 1st-order phase transition, and it is a direct consequence of the strong Ising anisotropy of the magnetic moments.

\begin{figure}[tb]
\includegraphics[width=8.5cm]{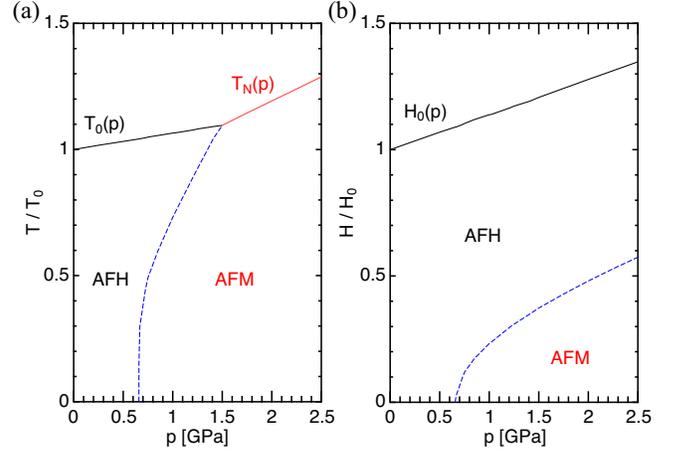}
\caption{(color online) The MF phase diagrams, (a) $p$-$T$ and (b) $p$-$H$. The temperature and the $c$-axis magnetic field are scaled by those of the MF critical values at ambient pressure.}
\label{phase}
\end{figure}

With the same model parameters, we discuss the magnetic-field dependence along the c-axis.
The increase of $H$ continuously decreases the intensity relatively faster at ${\bm Q}$ than at ${\bm Q}^{*}$ as shown in Fig.~\ref{skw}(c) and (d).
Note that the dispersion of the shadow band due to the antiferro holding gains a slight intensity at $H/H_{0}=0.5$.
With further increase of $H$, the spectra in the high-$H$ paramagnetic phase (not shown) is similar to those of Fig.~\ref{skw}(b) in the AFH phase.
The tendency of the field dependence is consistent with the experiments \cite{Bourdarot03,Santini00}.
The obtained MF phase diagrams are summarized in Fig.~\ref{phase}, where $H$ and $T$ are scaled by the critical values of the AFH phase at ambient pressure, $H_{0}$ and $T_{0}$.

Now, let us examine the magnetic-exciton mechanism of the superconductivity.
We consider the effective ``duality'' coupling between the heavily renormalized quasiparticles and the magnetic exciton as
\begin{equation}
H_{\rm int}=-g^{2}\sum_{\bm q}\chi_{z}({\bm q})s^{z}_{\bm q}s^{z}_{-{\bm q}},
\label{pairint}
\end{equation}
where $\chi_{z}({\bm q})={\rm Re}\,\chi_{z}({\bm q},0)$, $s_{\bm q}^{z}=\sum_{\alpha}^{\pm}\sum_{\bm k}\alpha\,c^{\dagger}_{{\bm k}\alpha}c^{}_{{\bm k}+{\bm q}\alpha}$ is the $z$-component of the quasiparticle spin density operator, and $g$ is the coupling constant between the quasiparticle and the magnetic exciton.
As the Cooper pair is formed by using the attraction mainly in the vicinity of ${\bm Q}$, we decompose $\chi_{z}({\bm q})$ into the form,
\begin{equation}
\chi_{z}({\bm k}-{\bm k}')=\chi_{z}({\bm Q})+\chi_{0}+\sum_{\gamma}\chi_{\gamma}f_{\gamma}({\bm k})f_{\gamma}({\bm k}'),
\end{equation}
where $\gamma=s,x,y,z,xy,yz,zx,xyz$, and the basis functions of the irreducible representation are defined as
\begin{subequations}
\begin{align}
&A_{1g}:\,\,\,f_{s}({\bm k})=c_{x}c_{y}c_{z},
\\
&B_{2g}:\,\,\,f_{xy}({\bm k})=s_{x}s_{y}c_{z},
\\
&E_{g}:\,\,\,f_{yz}({\bm k})=c_{x}s_{y}s_{z},
\,\,\,
f_{zx}({\bm k})=s_{x}c_{y}s_{z},
\\
&A_{2u}:\,\,\,f_{z}({\bm k})=c_{x}c_{y}s_{z},
\\
&B_{1u}:\,\,\,f_{xyz}({\bm k})=s_{x}s_{y}s_{z},
\\
&E_{u}:\,\,\,f_{x}({\bm k})=s_{x}c_{y}c_{z},
\,\,\,
f_{y}({\bm k})=c_{x}s_{y}c_{z}.
\end{align}
\end{subequations}
Here, $c_{\alpha}=\cos(\pi k_{\alpha})$ and $s_{\alpha}=\sin(\pi k_{\alpha})$, and the irreducible representations for the odd-parity pairing represent the symmetry of the orbital part.
The coefficients, $\chi_{\gamma}$, are obtained as
\begin{equation}
\chi_{\gamma}=\int_{0}^{1} \frac{d{\bm k}}{8}\int_{0}^{1}\frac{d{\bm k}'}{8}f_{\gamma}({\bm k})[\chi_{z}({\bm k}-{\bm k}')-\chi_{z}({\bm Q})-\chi_{0}]f_{\gamma}({\bm k}').
\end{equation}
By these decomposition, the pairing interactions for each channel $\gamma$ are given by $V_{\gamma}=g^{2}\chi_{\gamma}$ and $V_{s}\simeq g^{2}(\chi_{z}({\bm Q})+\chi_{0}+\chi_{s})$.
In the singlet channel and the triplet channel with ${\bm d}({\bm k})\,\|\,{\bm z}$ (${\bm d}({\bm k})\perp{\bm z}$), the negative (positive) sign means attractive because of the Ising form of the interaction, (\ref{pairint}).
The maximum attractive interaction gives the highest $T_{c}$.

\begin{figure}[tb]
\includegraphics[width=8.5cm]{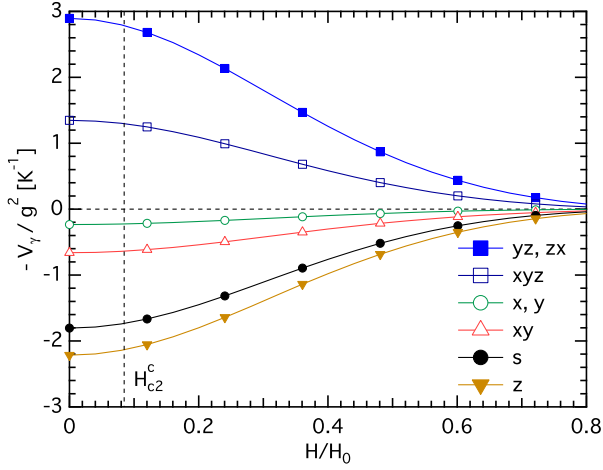}
\caption{(color online) The c-axis field dependence of the pairing interaction in each symmetry channel. In the singlet channel and the triplet channel with ${\bm d}({\bm k})\,\|\,c$ (${\bm d}({\bm k})\perp c$), the positive (negative) region means attractive. The $E_{g}$ ($yz$, $zx$) channel is always dominant in the whole $H$ region. The experimental $H_{c2}^{c}$ is indicated by the vertical dashed line.}
\label{att}
\end{figure}

The result of the decomposition under $H\,\|\,[001]$ is shown in Fig.~\ref{att}.
In the whole region of $H$, the maximum attractive interaction is obtained in the $E_{g}$ symmetry of ($yz$, $zx$) type.
The degeneracy of an arbitrary linear combination of the two-dimensional order parameter in $E_{g}$ at $T=T_{c}$ will be lifted below $T_{c}$.
The stable linear combination depends on microscopic details such as the topology of the Fermi surface.
From the pure energetics of the condensation energy for the isotropic Fermi surface, the chiral pairing is the most favorable \cite{Sigrist91},
\begin{multline}
f_{yz}({\bm k})+if_{zx}({\bm k})\propto
\\
\sin\frac{k_{z}c}{2}\left[\sin\frac{(k_{x}+k_{y})a}{2}+i\sin\frac{(k_{x}-k_{y})a}{2}\right],
\end{multline}
which has the line nodes in the basal plane and the point nodes along the c-axis.
In real space, the Cooper pair is formed between the nearest-neighbor interlayers.
This pairing symmetry is indeed proposed by the thermal transport and the heat capacity measurements \cite{Kasahara07,Kasahara09,Yano08}.


Let us examine the anisotropy in the magnetic field dependence of the pairing interaction.
Owing to the Ising-like field response of the magnetic exciton, the pairing interaction is almost independent of the field in the basal plane, while it is weakened by the c-axis field as shown in Fig.~\ref{att}.
It is the origin of the strong anisotropy in $H_{c2}$ as shown below. 

According to the Pesch approximation, which can be used to determine $H_{c2}(T)$ due to the orbital contribution \cite{Kusunose04}, the condition of $H_{c2}$ is given by
\begin{equation}
\ln\left(\frac{T}{T_{c}}\right)+\frac{1}{\lambda}\left(\frac{V_{\gamma}(0)}{V_{\gamma}(H)}-1\right)+2\pi T\sum_{n=0}^{\infty}\frac{1-I_{n}(T,H)}{\omega_{n}}=0,
\label{hc2eq}
\end{equation}
where $\lambda=\rho_{\rm F}|V_{\gamma}(0)|$ and $T_{c}=(2\omega_{c}e^{\gamma}/\pi)e^{-1/\lambda}$ are the dimensionless pairing interaction and $T_{c}$ at $H=0$.
In the case of ${\bm H}\,\|\,c$, the field dependence of the interaction is taken into account in the second term, while it can be neglected for ${\bm H}\perp c$.
$\omega_{n}=\pi T(2n+1)$ is the fermionic Matsubara frequency, and
\begin{equation}
I_{n}(T,H)=\sqrt{\pi}\langle u_{n}W(iu_{n})|\varphi_{\gamma}(\hat{\bm k})|^{2}\rangle,
\quad
u_{n}=\frac{2\Lambda\,\omega_{n}}{v_{\rm F}\sin\theta},
\end{equation}
with the Faddeeva function, $W(z)=e^{-z^{2}}{\rm erfc}(-iz)$ and the magnetic length, $\Lambda=(2|e|H)^{-1/2}$.
The bracket represents the angular average over the Fermi surface, where $\theta$ is the polar angle from the c-axis.
We have used $\varphi(\hat{\bm k})=\sqrt{15/4}\sin(2\theta)\sin\phi$ corresponding to the $E_{g}$ symmetry, and the isotropic Fermi surface with the velocity $v_{\rm F}$ for simplicity.

\begin{figure}[tb]
\includegraphics[width=8.5cm]{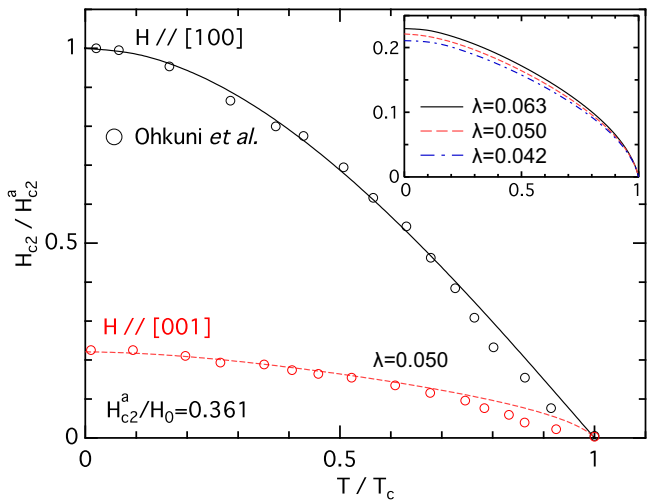}
\caption{(color online) The calculated $H_{c2}$ anisotropy. For ${\bm H}\,\|\,c$, the reduction of the pairing interaction as in Fig.~\ref{att} is taken into account. With the moderate coupling constant, $\lambda$, can reproduce the observed anisotropy indicated by the symbols \cite{Ohkuni99}. The inset shows the $\lambda$ dependence of $H_{c2}$ for ${\bm H}\,\|\,[001]$.}
\label{hc2}
\end{figure}

Figure \ref{hc2} shows the calculated $H_{c2}$ anisotropy with the experimental ratio, $H_{c2}^{a}/H_{0}=0.361$ \cite{Ohkuni99}.
The reduction of the pairing interaction $\lambda$ by $H$ decreases not only the absolute values of $H_{c2}$ and $T_{c}$ ($H=0$), but also enhances the $H_{c2}$ anisotropy as in (\ref{hc2eq}).
The inset of Fig.~\ref{hc2} shows the enhanced anisotropy due to the reduction of $\lambda$.
With the moderate coupling constant, $\lambda=0.05$, can reproduce the observed anisotropy indicated by the symbols \cite{Ohkuni99}.

Ordinary itinerant spin fluctuation is weakly influenced by the magnetic field because of its itinerant nature and relatively high energy scale.
On the contrary, the magnetic exciton is inherent in the localized character, and it is rather sensitive to magnetic fields, reflecting easily in the superconducting stability via the pairing interaction.
The feedback effect due the superconducting gap opening may be weak on the massive magnetic exciton spectra \cite{Bourdarot10}.
In reality, URu$_{2}$Si$_{2}$ is the low-carrier compensated metal in the HO phase, and hence the multi-gap pairing has been discussed \cite{Kasahara07,Kasahara09,Yano08}.
The extension of the present analysis to the multi-gap pairing would not alter the pairing symmetry and the qualitative aspect, but improve the precise form of the gap structure.
Due to a lack of the detailed information on the Fermi surfaces, it is beyond the present work.

In summary, we have addressed the magnetic exciton mediated superconductivity, which presumably emerges in the exotic AF hexadecapole ordered phase of URu$_{2}$Si$_{2}$.
Charactering the observed magnetic-exciton mode as the massive collective excitation from the AF hexadecapole ground state, we have deduced that the most favorable pairing is the chiral $d$-wave singlet, $k_{z}(k_{x}+ik_{y})$, in the $E_{g}$ symmetry with both the line and the point nodes.
Thereby, the Ising-like anisotropic field response of the mode results in the anisotropic reduction of the pairing interaction, yielding the strong anisotropy in $H_{c2}$.
The present picture based on the AF hexadecapole ordering provides a coherent description of the enigmatic property of URu$_{2}$Si$_{2}$, and it will shed light on further developments.

We would like to thank Masashige Matsumoto, Kazuo Ueda, Kazumasa Miyake, Collin Broholm, Hisatomo Harima, Koichi Izawa, Dai Aoki, Hideki Tou, Fr\'ed\'eric Bourdarot, and Yuji Matsuda for fruitful discussions.
This work was supported by a Grant-in-Aid for Scientific Research on Innovative Areas ``Heavy Electrons" (No.20102008) of The Ministry of Education, Culture, Sports, Science, and Technology (MEXT), Japan.


\end{document}